\documentclass[10pt,prl,twocolumn,superscriptaddress]{revtex4}

\usepackage{graphicx}
\usepackage{amssymb}
\usepackage{times}
\usepackage{amsmath}
\usepackage{amsfonts}
\usepackage{txfonts}
\usepackage{color}

\begin{document}

\title{Near threshold all-optical backaction amplifier}

\author{Terry G. McRae} \affiliation{Centre for Engineered Quantum Systems, School of Mathematics and Physics, University of Queensland, St Lucia, Brisbane, QLD 4072, Australia} %\affiliation{MacDiarmid Institute, Physics Department, University of Otago, Dunedin, New Zealand}

\author{Warwick P. Bowen} \affiliation{Centre for Engineered Quantum Systems, School of Mathematics and Physics, University of Queensland, St Lucia, Brisbane, QLD 4072, Australia}

\begin{abstract}
A near threshold all-optical backaction amplifier is realized. Operating near threshold in an integrated micron-scale architecture allows a nearly three orders of magnitude improvement in both gain and optical power requirements over the only previous all-optical implementation, with 37~dB of gain achieved for only 12 $\mu$W of input power. Minor adjustments to parameters allows optical filtering with narrow bandwidth dictated by the mechanical quality factor. Operation at cryogenic temperatures may enable standard quantum limit surpassing measurements and ponderomotive squeezing.
\end{abstract}

\date{\today} \maketitle
%\pacs{03.67.Hk  03.67.Dd  42.65.Lm}

Optical amplifiers are ubiquitous in present-day communications and sensing technologies, and are fundamental to many high precision scientific endeavors. The classic example is the erbium doped fiber amplifier (EDFA) which, with pump power and gain of around 100 mW and 20 dB respectively\cite{DesurvireBeckerEDFA}, dramatically reduces power consumption and speeds up performance in long-haul fiber optic communications, enabling modern optical telecommunications networks. It is well known that phase sensitive amplifiers based on the optical parametric nonlinearity offer the means to both further reduce pump power, with nonlinear thresholds at the level of microwatts\cite{KippenbergVahalaToroidOPO,FurstLeuchsOPO}, and boost noise performance past the 3 dB quantum limited noise figure of phase insensitive amplifiers\cite{TongGrunerNielsenPSA11}. This letter reports an on-chip all-optical phase sensitive amplifier where, in contrast, the nonlinearity arises, not from an intrinsic optical nonlinearity, but rather from the radiation pressure driven dynamical interplay between the optical and mechanical structure of a microcavity optomechanical system.
% in a complimentary fashion to measurement noise reduction using squeezed states of light \cite{XiaoKimbleSqueezedDetection1987,GrangierLaPortaSqueezedDetection1987,McKenzieMcClellandSqueezedDetection2002}.
%
Such {\it backaction amplifiers}\cite{Verlot10} are but one example of the rapid progress currently occurring in the emergent field of cavity optomechanics based on the dynamical backaction between light and matter; with recent demonstrations of mechanical cooling\cite{ArcizetHeidmannCooling2006,GiganZeilingerCooling2006,SchliesserKippenbergCooling2006}, optomechanical induced transparency (OMIT)\cite{WeisKippenbergOMIT10}, regenerative mechanical oscillation\cite{KippenbergVahalaRadiationPressure05,RokhsariVahalaOscillator05,ArcizetHeidmannCooling2006},
%measurement of nanomechanical motion with an imprecision below the standard quantum limit (SQL)\cite{AnetsbergerKippenbergMeasurementSQL10},
optomechanical induced chaotic oscillation\cite{CarmonVahalaOptomechanicalChaos07}, tunable phonon lasers\cite{GrudininVahalaPhononLaser2010} and even technological applications such as a photonic radio frequency (RF) down converter\cite{HosseinVahalaPhotonicMixer}.

The first demonstration of backaction amplification, and only previous all-optical demonstration, used an electronically locked Fabry-Perot cavity with milligram effective mass\cite{Verlot10}. However, due to the close proximity of parametric instability, operation was precluded near the threshold for optomechanical parametric oscillation where the amplification is strongly enhanced.
%
%The first experimental demonstration of %all-optical
%backaction amplification used an electronically locked Fabry Perot cavity\cite{Verlot10}, with the proximity of parametric instability precluding operation near the threshold for optomechanical regenerative oscillation where the amplification is significantly enhanced.
%Instead the more stable cooling (red detuned) side of optical resonance was used.
This
%factor, combined with the comparatively high mass of the vibrating mirror (72 mg) and relatively low optical quality factor ($7.6\times10^5$)
limited the gain achieved to 8~dB, with 10~mW of input optical power. Here, by contrast, the optomechanical resonator is an integrated monolithic silica microtoroid\cite{Armani03} with microgram effective mass, high mechanical stability, and ultrahigh optical quality factor. Critically, the thermal bistability\cite{RokhsariVahalaBistability2004} of silica provides strong thermo-optic locking to the blue detuned side of optical resonance\cite{CarmonVahalaThermalStability2004,McRaeBowenThermalLock2009}. This allowed operation close to parametric instability, resulting in a nearly three orders of magnitude improvement in both gain and optical power requirements, with 37dB of gain achieved for only 12~$\mu$W of input power.

%By operating close to the threshold for parametric instability 37dB of gain is achieved with only 12 $\mu$W of input power.

\begin{figure}[t!]
\begin{center}
\includegraphics[width=8cm]{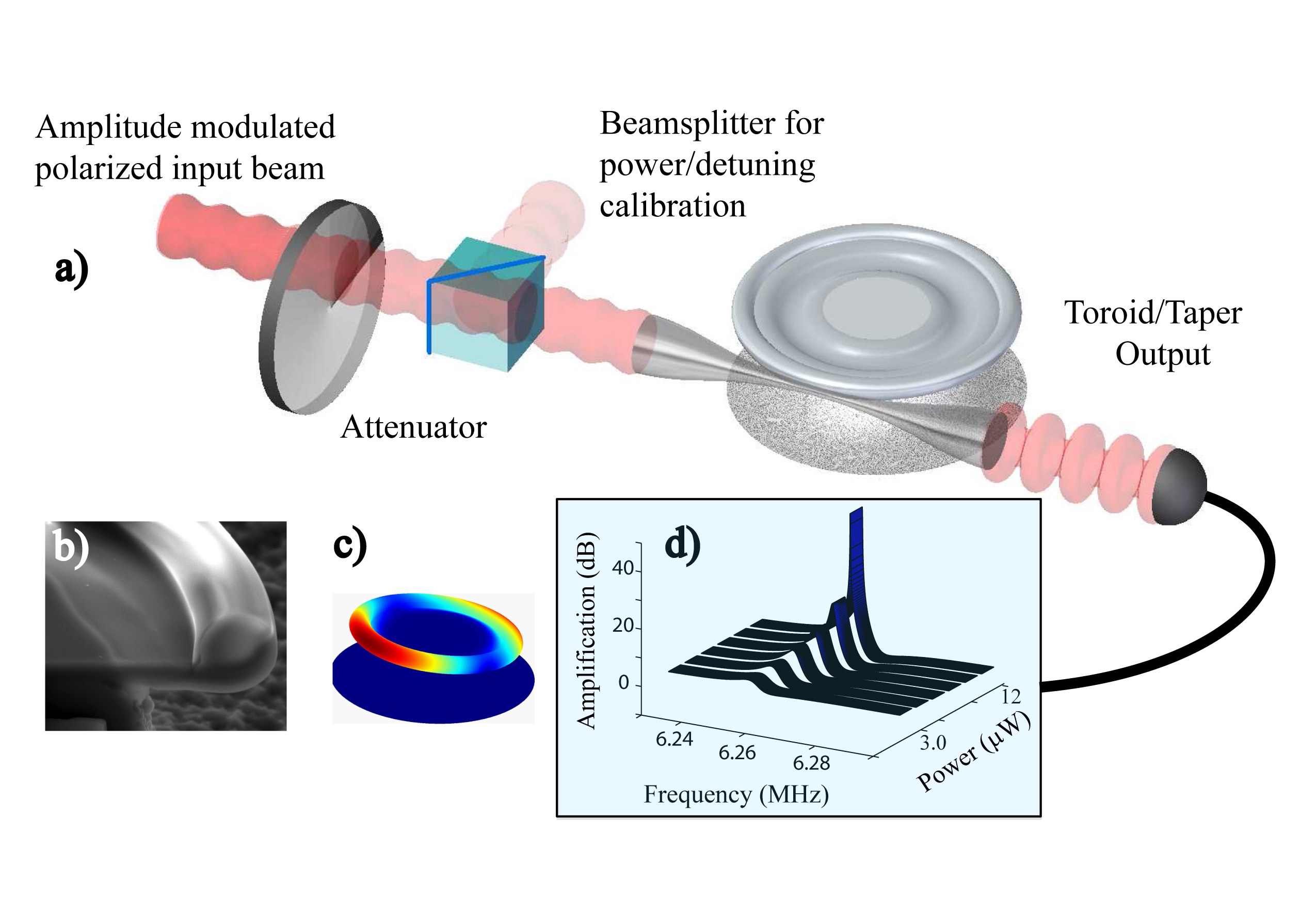}
\caption{(color online) a) Illustration of the backaction amplification experiment.  b) SEM image of a microtoroid cut via focused ion beam milling to show the rim offset relative to the plane of the disk. The rim  diameter was characterised to be $4.5\pm 0.5 \mu$m. c) Finite element model of the mechanical first order crown mode used for this work. d) Typical optomechanical response determined via network analysis showing amplification increasing as the incident power approaches threshold.}
\label{schematic}
\end{center}
%\vspace{-8mm}
\end{figure}

Parallel to the all-optical architecture reported here, near threshold backaction amplification has very recently been reported in two other architectures, microwave signal amplification via microwave resonators capacitively coupled to a micromechanical resonator\cite{Massel2011}, and optical signal amplification and ponderomotive squeezing via the coupling of the centre-of-mass-motion of a cloud of ultracold atoms to a Fabry-Perot cavity field\cite{BotterStamperKurnOMsqueezing2011}. In the latter case, the atto-gram mass of the atom cloud led to remarkably low power requirements, with 20 dB of gain observed for tens of pico-Watts of input power. In contrast to these developments, our architecture is compatible with radiation pressure driven on-chip photonic circuits and photonic communication\cite{Tang}, with fabrication achieved using standard lithography techniques suitable for integration into wafer scale technologies and the capacity for highly efficient direct coupling to telecommunications optical fiber. Furthermore, when cooled into a cryogenic regime, the quantum limited signal amplification provided by this system could enable length fluctuations of the cavity to be measured with sensitivity far surpassing the standard quantum limit (SQL)\cite{ArcizetPinardBeatingSQL2006}.

Photon reflection at the surface of an optical resonator transfers momentum to its mechanical structure with the modified resonator geometry in turn acting back upon the radiation field\cite{KippenbergOptMechReview}. This radiation pressure induced mutual coupling of the optical and mechanical modes has been considered theoretically for decades\cite{Braginsky,CavesQMNoise1981}, particularly in the context of gravitational wave interferometers\cite{BraginskyVyatchaninParametricInstability2001}. In addition to imposing the SQL as a fundamental detection limit, the static backaction of the light on the cavity can influence the behavior of the optomechanical resonator, including such effects as optical bistability\cite{DorselWaltherBistability1983} and radiation pressure induced changes in the mechanical rigidity of the cavity, referred to as the optical spring effect\cite{SheardWhitcombOpticalSpring2004}. The observation of dynamic backaction, where the decay time of photons inside the cavity is comparable to or longer than the mechanical oscillator period, has recently become possible with advances in nano and micro fabrication techniques.

Detuning the optical field incident on a cavity optomechanical system onto the blue side of resonance causes energy transfer from the optical field into the mechanical oscillator\cite{KippenbergOptExp}. At sufficiently high optical power a threshold condition occurs where mechanical loss is completely offset by mechanical gain. Beyond this power level, regenerative mechanical oscillation occurs with exponential growth in the mechanical oscillation followed by inevitable saturation. Key to the optomechanical amplifier reported here is to operate just below threshold where mechanical fluctuations are greatly amplified, without the presence of saturation. In this regime, amplitude modulations on the optical field entering the microresonator are imprinted via radiation pressure onto the mechanical motion, amplified by the optomechanical interaction, and then encoded back onto the phase of the intracavity field. Upon exiting the cavity, depending on the optical detuning and coupling rate constructive or destructive interference can be achieved with the fluctuations reflected directly from the cavity. This enables dynamical backaction amplification\cite{Verlot10}, narrowband spectral filtering, and ultimately ponderomotive squeezing\cite{BotterStamperKurnOMsqueezing2011}.

Figure \ref{schematic}(A) illustrates the experimental setup of the backaction amplifier reported here. A New Focus Velocity laser was tuned, polarized, amplitude modulated and coupled via tapered optical fiber to a silica microtoroid\cite{Armani03}.  The input power was controlled by an optical attenuator, and the amplitude modulation was provided by a network analyzer (Agilent E5061A) signal applied to the input beam via a Mach Zehnder interferometer with depth of modulation chosen to ensure good contrast from mechanical Brownian noise for all input power levels. The input optical field was detuned to the blue side of optical resonance causing the desired regenerative amplification, and imprinting the cavity length fluctuations due to mechanical motion onto the amplitude quadrature of the output field. The output field was directly detected on a silicon photodiode (New Focus 1801-FC). Throughout data taking it was critical that both optical detuning and coupling rate were maintained stably. Both parameters were dependent on the incident optical power. As the incident power is increased the detuning decreases due to enhanced thermal locking, and the optical coupling rate increases due to the enhanced cavity evanescent field which increases the dipole force on the taper and pulls it into closer proximity to the toroid. Consequently, techniques were developed to monitor and stabilise both parameters in real time. The optical detuning was controlled throughout the experiments by tapping off a small fraction of the input and output optical fields and adjusting the frequency of the laser to maintain the ratio at a constant level. The optical coupling rate was monitored by applying a dual frequency amplitude modulation to the input field, with one frequency (100 MHz) outside and the other (30 MHz) approximately equal to the optical linewidth. The phase of the output signal at 30 MHz was then much more sensitive to the optical coupling rate, than that of the 100 MHz signal. The ratio of the two signals, after being mixing down to DC, provided a control diagnostic and the taper position was adjusted to maintain this at a constant level throughout experiments. Both frequencies were far greater than the mechanical resonance frequency (6.254 MHz), and hence these additional modulations had no effect on the optomechanical response.

\begin{figure}
\begin{center}
\includegraphics[width=8cm]{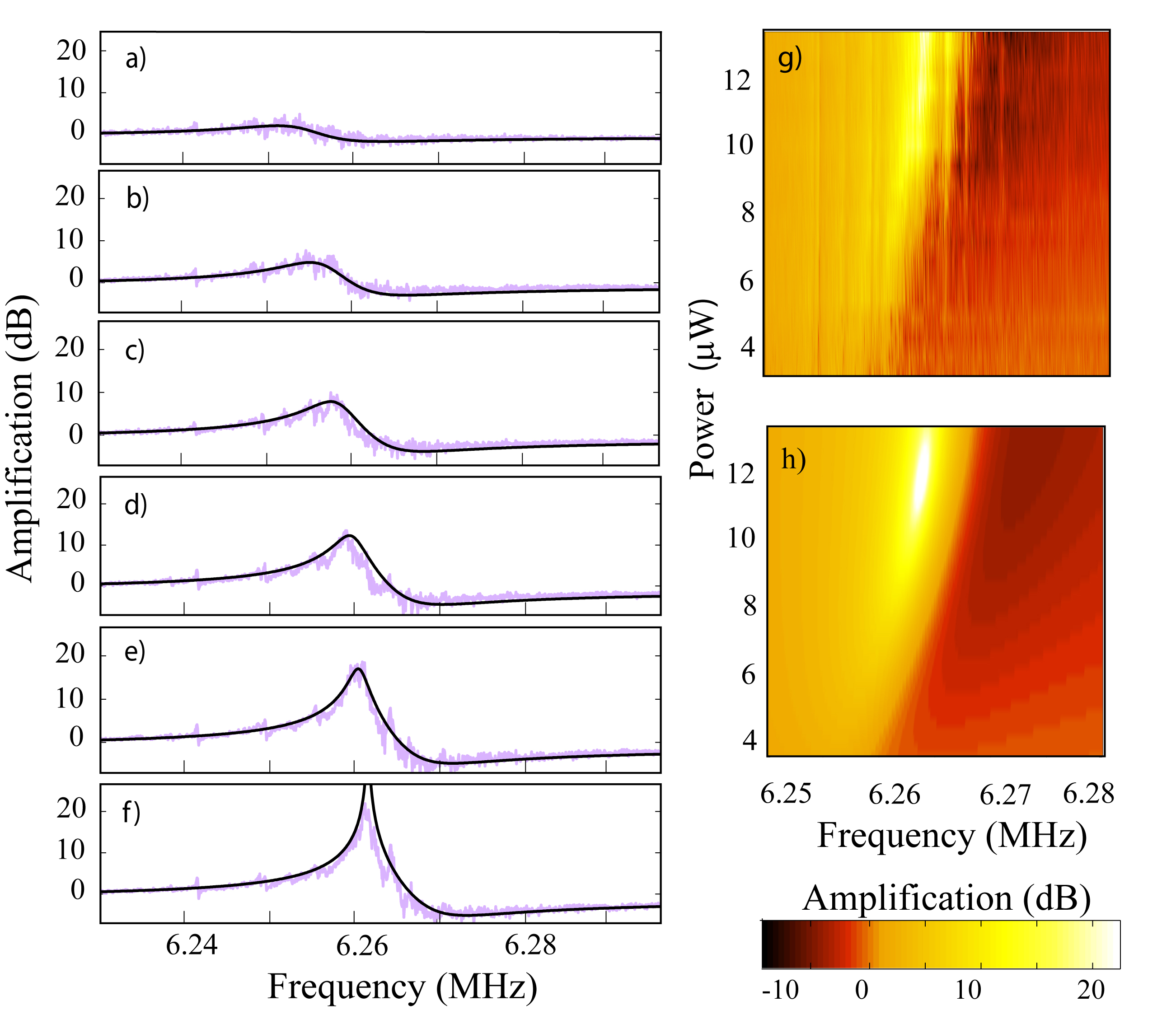}
\caption{(color online) Backaction amplification as a function of frequency for various pump powers. Network analyzer bandwidth: 300 Hz, sweep time: 5 seconds. Grey traces: experimental results. Black lines: model. Incident power in $\mu$W: a) 3.0, b) 5.4, c) 7.2, d) 9.0, e) 10.2, f) 12.0. Figure g): Optomechanical response for transmitted power versus input power and frequency. Figure h): Model of optomechanical response versus input power and frequency.  Model parameters: $Q_0=4\times10^7$, $Q_m=480$, $m_{\rm{eff}}=64 \mu g$, optical detuning $\Delta=3$ MHz, input coupling rate $\gamma_{\rm in}/2 \pi = 13$~MHz, strength of Kerr nonlinearity $\Lambda/2\pi=0.30+0.14i$ MHz (supplementary information).}
\label{IntensityResponse}
\end{center}
\end{figure}

The mechanical noise spectrum was determined via spectral analysis of the output photocurrent without applied amplitude modulation. As the input power was increased a gradual increase in both the mechanical quality factor and mechanical noise power was observed, as expected in the approach to regenerative amplification. The onset of regenerative amplification was found to occur at an input power of $12~\mu$W at which point the balancing of dissipation with gain resulted in a dramatic increase in mechanical noise power as expected, and consistent with previous work\cite{KippenbergVahalaRadiationPressure05}. Applying the amplitude modulation and sweeping its frequency across the mechanical resonance allowed the optomechanical response to be characterized via network analysis. Fig.~\ref{IntensityResponse}(a-f) show the intensity response, and associated theoretical model (supplementary information) as the incident power approaches regenerative amplification threshold. At frequencies lower than the mechanical resonance frequency optical amplification is observed due to optomechanical amplification within the resonator, and subsequent constructive interference between the light exiting the resonator and that transmitted directly through the fiber. Above the mechanical resonance frequency, by contrast, deamplification occurs due to the phase change in the mechanical response, and the resulting destructive interference. As the pump power is increased towards threshold with the detuning and coupling rate kept constant the magnitude of the amplification increases and the mechanical resonance frequency is shifted due to the optical spring effect. Fig.~\ref{IntensityResponse}(g) and Fig.~\ref{IntensityResponse}(h) compare the theoretically predicted and experimentally observed optomechanical response for eighteen different power levels, with excellent agreement obtained once the intensity dependent phase shift due to the optical Kerr non-linearity of silica is taken into account (supplementary information).

\begin{figure}
\begin{center}
\includegraphics[width=8cm]{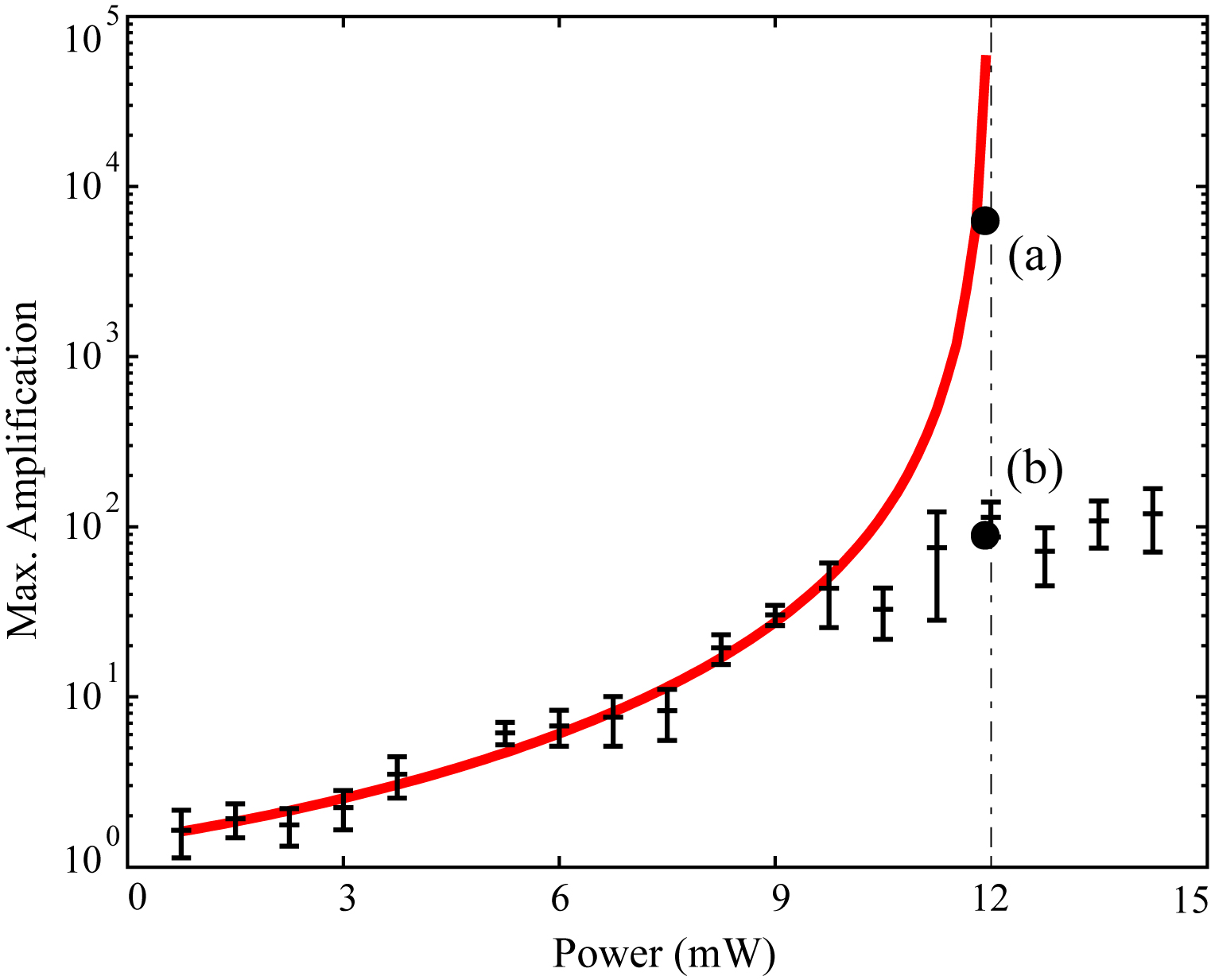}
\caption{Maximum amplification as function of input power. Points (a) and (b), comparison of amplification achieved with (a) and without (b) reducing the input modulation by a factor of 24 dB. Solid line: blue detuned model.
Dotted line: Equivalent predictions for red detuning.
Vertical line indicates threshold. Model parameters same as for Fig.\ref{IntensityResponse}.}
\label{PwrCurve}
\end{center}
\end{figure}
The peak mechanical amplification as a function of power is shown in Fig.~\ref{PwrCurve}. It can be seen that the amplification increases close the threshold in good agreement with theory, but saturates
%at optical powers around 10 $\mu$W,
at a maximum of around 22 dB. This saturation was due to the depth of modulation on the output field approaching 100~\%,
%as confirmed by direct time domain measurements,
and was eliminated by reducing the input depth of modulation by 24 dB with other experimental conditions unchanged. This allowed a maximum of 37 dB of amplification to be achieved in good agreement with theory, as shown by the two additional measurements labeled (a) and (b) in Fig.~\ref{PwrCurve}.
%
% two additional measurements were made labeled (a) and (b) in Fig.~\ref{PwrCurve}. Experimental conditions were unchanged, with the exception that for point (a) the input amplitude modulation was reduced by 24 dB allowing a maximum of 37 dB of amplification to be achieved in good agreement with theory.
%This saturation was eliminated by reducing the input amplitude by \textbf{XXXX} dB as shown by the experimental point labelled (a) in Fig.~\ref{PwrCurve}, allowing a maximum of 37 dB of amplification to be achieved.
%For comparison the dashed line shows the gain achievable when the optical field is red detuned, which is more than three orders of magnitude below what is achieved here.

\begin{figure}
\begin{center}
\includegraphics[width=8cm]{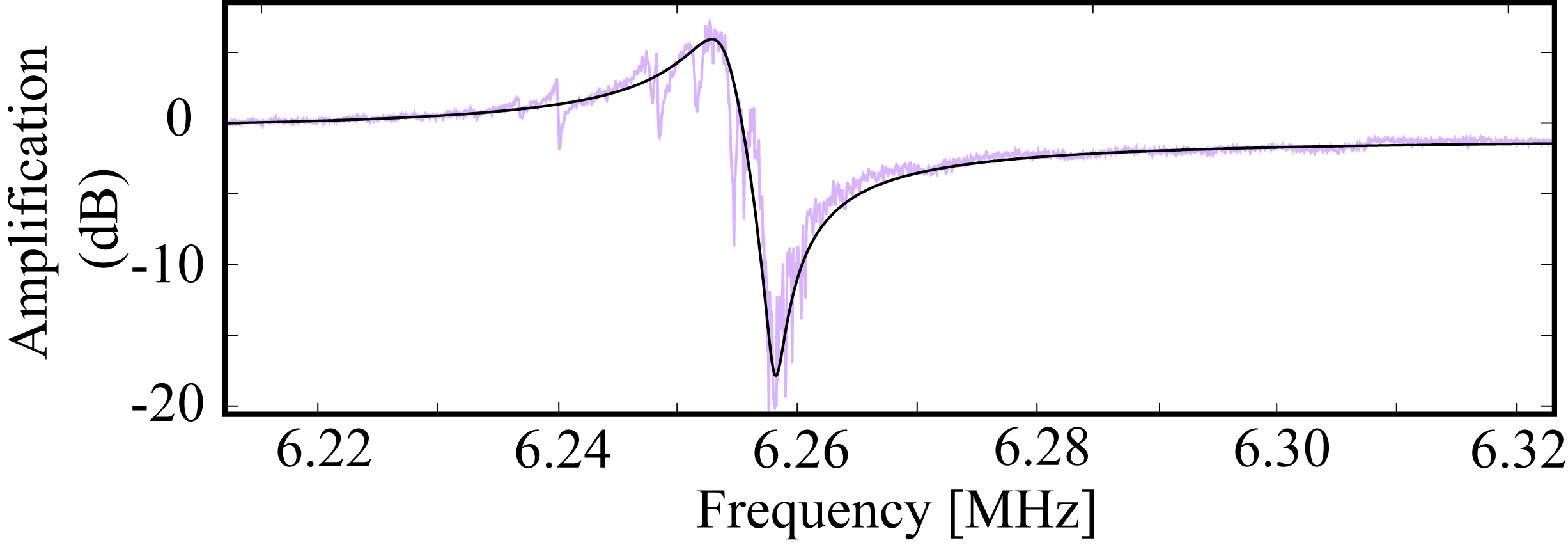}
\caption{Optomechanical response with parameters optimized for destructive interference. Black line: model. Grey trace: experiment. Optical power at approximately 60~\% of threshold, $Q_0=6\times10^7$, $Q_m=480$, $m_{\rm{eff}}=64 ~\mu$g, $\Delta=2.92$ MHz, input coupling rate $\gamma_{\rm in}/2 \pi = 8$~MHz, strength of Kerr nonlinearity $\Lambda/2\pi=0.30+0.14i$ MHz (supplementary information).}
\label{Notch}
\end{center}
\end{figure}
By optimizing the destructive interference between the directly transmitted and optomechanically amplified modulations,  the backaction amplifier demonstrated here can also be used as a narrow band all optical filter and, at cryogenic temperatures, to generate ponderomotive optical squeezing\cite{BotterStamperKurnOMsqueezing2011}. We demonstrate the former here by undercoupling the optical field and operating well below threshold.
Figure~\ref{Notch} shows the typical optical response
%of the system
in this regime with a narrow band signal suppression of around 20 dB with a 3~kHz  full width half maximum.
%The filter frequency can be tuned by taking advantage of the optical spring effect, or by applying an electro-optic spring as is common in the gravity wave community\cite{???}\textbf{[REF]}; and more coarsely via mechanically engineering the physical structure.
As has been pointed out before\cite{BotterStamperKurnAmplifierOMIT2011}, this narrowband destructive interference has similarities to optomechanical induced transparency (OMIT) where the intracavity probe field destructively interferes when the two photon resonance condition is met\cite{WeisKippenbergOMIT10}.
%\textbf{This tunability in OMIT systems allows access to wavelength regions in the technologically important infrared region previously inaccessible to electromagnetically induced transparency. OMIT technology has recently enabled strong electromechanical coupling and was a precursor to ground state cooling of a macroscopic oscillator \cite{TeufelSimmondsStrongCoupling11,TeufelSimmondsGroundState11}. Double disk and "zipper" cavities have also exhibited OMIT and has the potential for the development of an on chip platform for information processing and storage either in the classical or quantum regime\cite{LinPainterZipperOMIT10}.}
%These results lay the foundation for future work where radiation pressure actuated devices may used in a variety of photonics applications\cite{RosenbergPainterWGMrouting09,WiederheckerLipsonPhotonicStructures09}.

%Our experiment was not performed in the cryogenic regime, or in vacuum, and therefore the sensitivity was limited by thermal noise on the resonator and is far from the SQL \textbf{[calculate how far]}.
\begin{figure}
\begin{center}
\includegraphics[width=8cm]{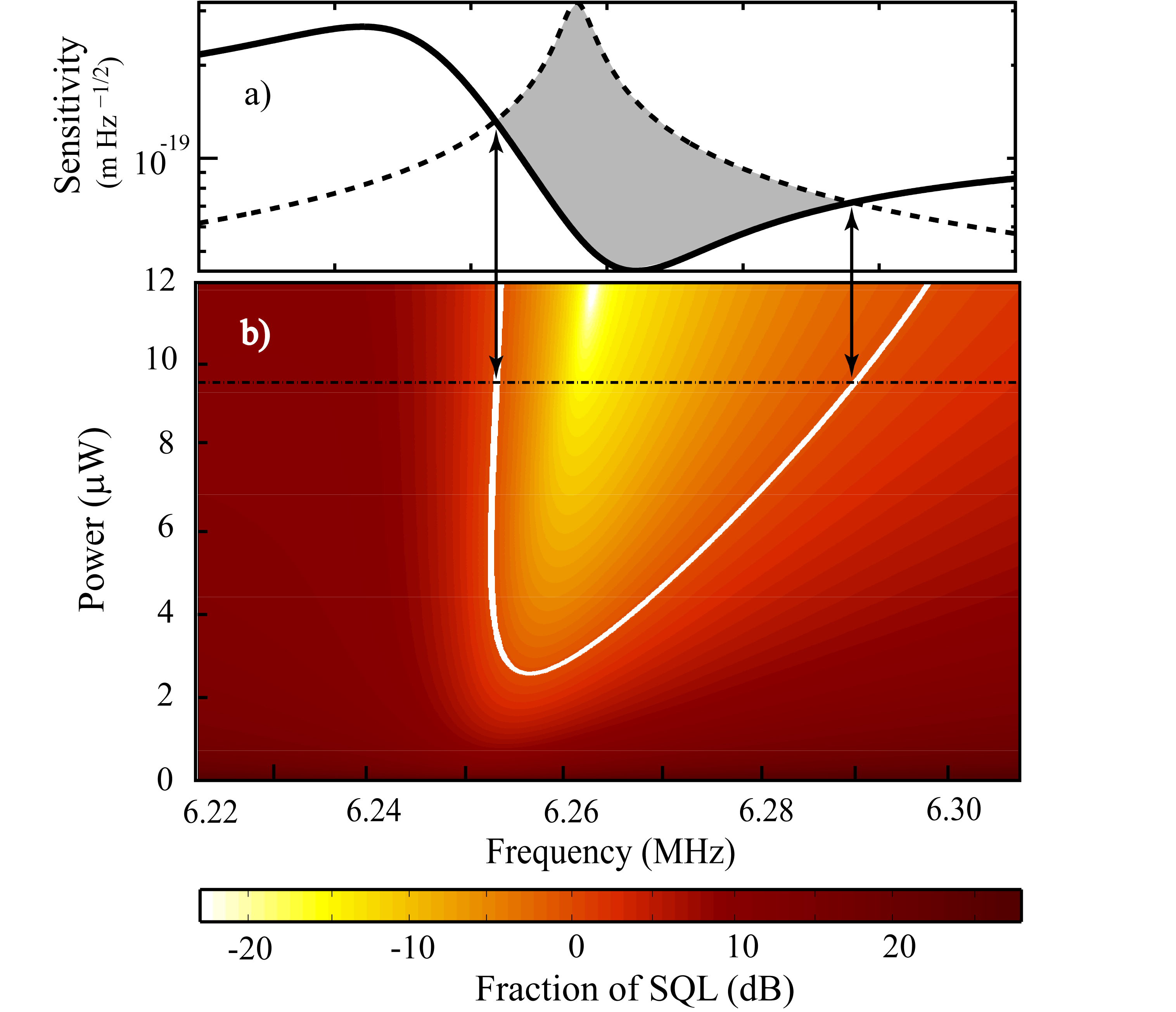}
\caption{(color online) Surpassing the SQL using dynamical backaction. a) Comparison of displacement sensitivity (solid line) to SQL (dotted line) for an input power of 9.6~$\mu$W and the optical and mechanical parameters of the optomechanical amplifier demonstrated here, where the SQL is given simply by $\hbar |\chi_{\rm eff}|$ with $\chi_{\rm eff}$ being the optically modified mechanical susceptibility\cite{ArcizetPinardBeatingSQL2006}. b) Ratio of displacement sensitivity and SQL as a function of frequency and incident power. Model parameters same as for Fig.\ref{IntensityResponse}.}
\label{SQL}
\end{center}
\end{figure}
By amplifying the intracavity phase modulation on an optical field, cryogenically cooled optomechanical amplifiers are predicted to be capable of displacement sensitivity beyond the SQL \cite{Verlot10,ArcizetPinardBeatingSQL2006} (see supplementary material). Fig.~\ref{SQL} shows the level of sensitivity enhancement predicted for the device reported here using our experimentally determined parameters (see supplementary information for calculation).
%Using the experimentally determined parameters for our optomechanical amplifier, it is possible to determine the level of sensitivity enhancement that should be possible with the device reported here.
Fig~\ref{SQL}(a) compares the predicted displacement sensitivity and SQL as function of frequency for an input power of 9.6 $\mu$W.
%, and the calculation of displacement sensitivity can be found in the supplementary information.
As can be seen, it should be possible to surpass the SQL over a significant frequency band near the mechanical resonance frequency. Fig. \ref{SQL}(b) shows the ratio of displacement sensitivity to SQL as a function of input power, predicting that close to regenerative amplification threshold the SQL may be surpassed by several orders of magnitude.
%This is in sharp contrast to techniques utilising optical squeezing to surpass the SQL, where even a factor of ten would be extremely challenging due to limitations in the level of achievable squeezing caused by optical inefficiencies\cite{VahlbruchSchnabel10dBSqueeze09}.

%\begin{figure}
%\begin{center}
%\includegraphics[width=8cm]{Fig4Pcolor.eps}
%\caption{.}
%\label{driveschematic}
%\end{center}
%\end{figure}

\emph{Conclusion} Near threshold all-optical backaction amplification has been demonstrated, driven by radiation pressure and achieving thirty seven decibels of amplification with only 12 $\mu$W of input power. With different operating conditions the same optomechanical system acts as a notch filter suppressing optical amplitude fluctuations over a narrow frequency range. This represents an enabling step towards future all-optical radiation pressure driven photonic circuits and communication systems, and the achievement of ponderomotive squeezing and light matter entanglement in integrated structures.

\emph{Acknowledgements} This research was funded by the Australian Research Council Centre of Excellence CE110001013 and Discovery Project DP0987146. Device fabrication was undertaken within the Queensland Node of the Australian Nanofabrication Facility. Special thanks to Kwan Lee for the FIB image in fig.~\ref{schematic}B.

\end{document}